\documentclass[epsf,twocolumn,showpacs,preprintnumbers]{revtex4}
\usepackage{graphics}
\usepackage{graphicx}
\usepackage{dcolumn} 
\usepackage{bm}
\usepackage{epsfig}
\begin{document}
\title{Disproportionation and Critical Interaction Strength 
   in Na$_x$CoO$_2$: Concentration Dependence}
\author{K.-W. Lee and W. E. Pickett} 
\affiliation{Department of Physics, University of California, 
     Davis, CA 95616}
\date{\today}
\begin{abstract}
We present results of studies of charge disproportionation (CD)
and spin differentiation in Na$_x$CoO$_2$ 
using the correlated band theory approach
(local-density approximation$+$Hubbard U:LDA+U).
The simultaneous CD and gap opening can be followed through a first order
charge disproportionation transition.
By comparison with experiments, we propose a value of
the Coulomb repulsion strength U($x$) that has
significant dependence on the carrier concentration $x$, for which we
obtain a phase diagram.  The connection between the oxygen height and
effects of on-site correlation is also reported.
\end{abstract}
\maketitle

Since the discovery of superconductivity with T$_c$ $\sim$ 4 K 
in water-intercalated Na$_x$CoO$_2$ (NxCO) by Takada et al.
in the early of 2003,\cite{takada} several hundred papers 
have been published.
The system shows several similarities to high T$_c$ superconducting 
cuprates: layered transition metal oxide; vicinity of (presumed)
Mott insulator CoO$_2$; variation with carrier concentration $x$;
a ``superconducting dome.''
However, it also has differences from cuprates: electron-doped system
from Mott insulator phase, actually nonmagnetic metallic 
sister $x=0$ phase, triangular lattice which can be frustrated,
different superconducting dome shape, and low T$_c$.

In addition to its interesting properties in view of superconductivity,
its normal state has a rich phase diagram depending on carrier
concentration $x$.\cite{foo}
For $x<0.5$ the Pauli-like susceptibility has been observed, while
for $x>0.5$ the susceptibility shows local moment (Curie-Weiss) behavior.
For $0.75 \le x \le 0.85$, NxCO shows antiferromagnetic (AFM) alignment 
of ferromagnetically ordered CoO$_6$ layers.\cite{x075}
Specifically at $x$=0.5, the system has been observed to undergo
a charge disproportionation (2Co$^{3.5+} \rightarrow$ Co$^{3+}$+Co$^{4+}$)
and metal-insulator transition at 50 K, while the rest of the phase diagram
is metallic.\cite{foo}

We have focused on normal state electronic and magnetic structure of NxCO,
with specific interest in ordering near the ground state, and characterizing
the strength and consequences of correlation effects. 
In transition metal oxides in general, the mechanism of charge ordering or
charge disproportionation (CD) is an important issue, and our work 
provides new inroads in the understanding of the mechanisms of
disproportionation.

 Our calculations were based on the supercell approach,\cite{prb04,prb05} 
using the  full-potential local-orbital method (FPLO).\cite{klaus}
Both popular schemes for LDA+U functional were monitored. 
(The intra-atomic exchange integral $J$=1 eV was left 
unchanged.) Both have the same Hubbard-like density-density 
interaction, but differ in just how to subtract out 
``double counting" of the pair interaction.  One way, 
the so-called Fully Localized 
Limit (or Atomic Limit),\cite{ldau1} 
is an atomic-like treatment and is appropriate
for large on-site Coulomb repulsion U, 
{\it e.g.} strong localization of the correlated orbital. 
Another is often called the Around Mean Field\cite{ldau2} and 
is more appropriate when U is not so strong. In this sodium cobaltate,
our calculations have shown similar results from both schemes.

\begin{figure}[tbp]
\rotatebox{-90}{\resizebox{5.9cm}{5.9cm}{\includegraphics{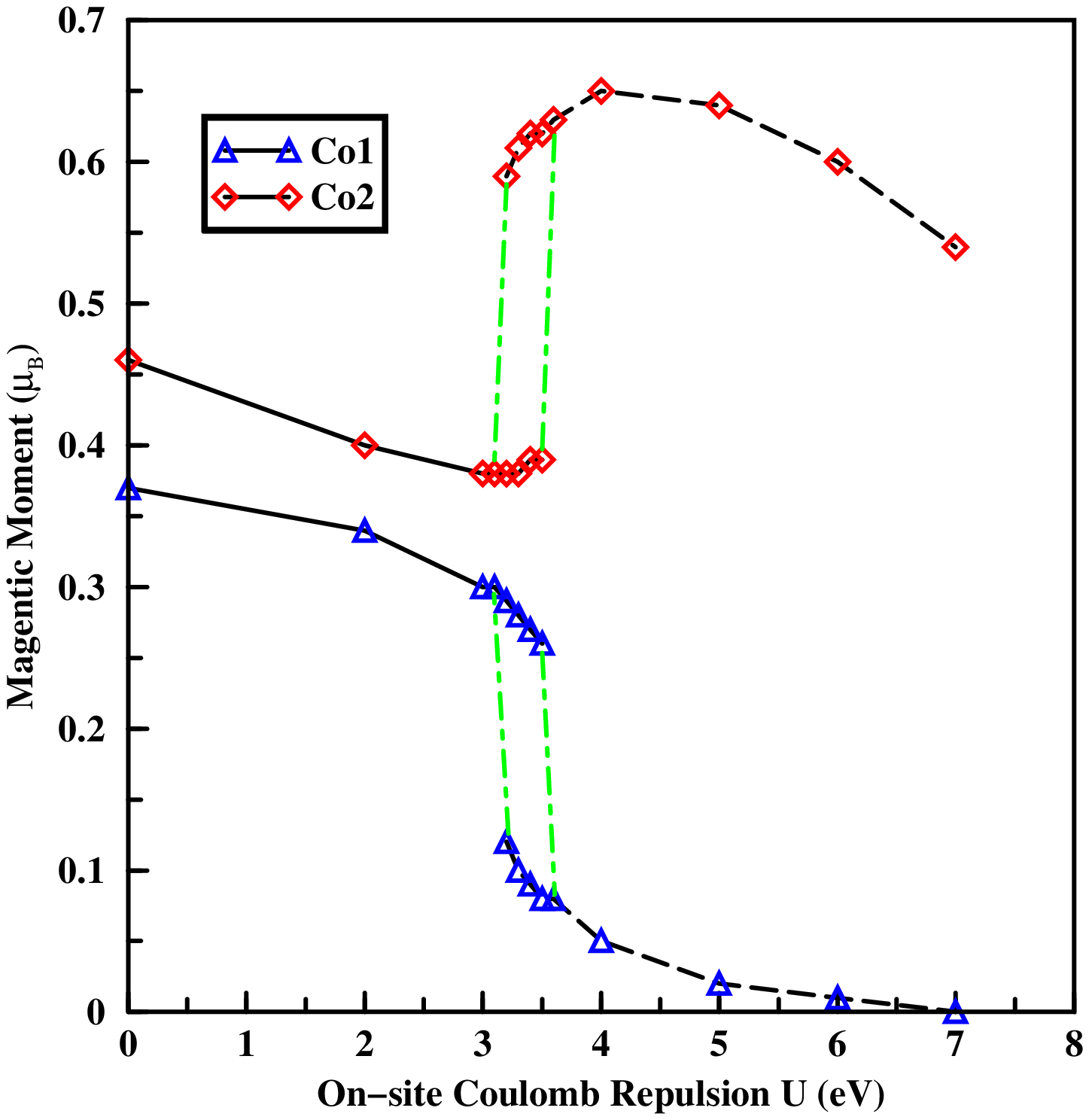}}}
\rotatebox{-90}{\resizebox{5.6cm}{7.0cm}{\includegraphics{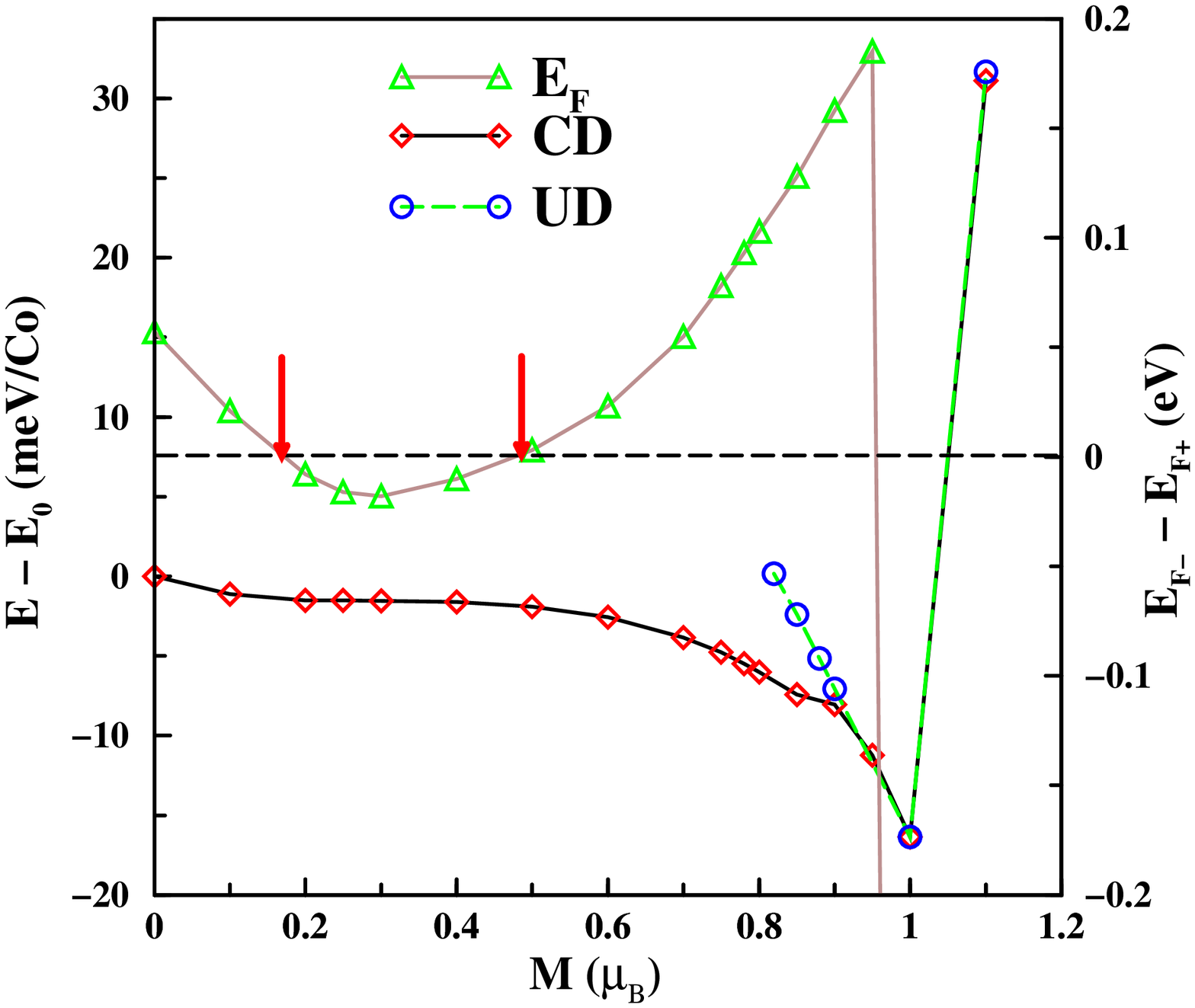}}}
\caption{Upper panel:
 Effect of the on-site Coulomb repulsion U on the magnetic
 moments for ferromagnetic ordering in Na$_{0.5}$CoO$_2$.
 The change shows a first-order transition (and accompanying
 hysteresis) in the critical region (3.2 eV to 3.6 eV).
 Metal-insulator transition simultaneously occurs with CD
 (Co1$\rightarrow$Co$^{3+}$, Co2$\rightarrow$Co$^{4+}$) due to nature
 of the first-order transition. Dashed and solid lines indicate
 the CD and UD states, respectively.
 Bottom panel:
 Fixed spin moment calculation of energy vs. total magnetic moment $M$
 (in $\mu_B$/2Co) at U=3.3 eV in Na$_{0.5}$CoO$_2$.
 The left and right sides of the y-axis denote the total energy
 difference and difference between minority and majority Fermi energy
 (effective applied field), respectively.
 The arrows mark zeroes of the Fermi energy difference, which correspond
to extrema of the energy.}
\label{um}
\end{figure}

Within the local density approximation (LDA), all Co magnetic moments
are nearly equivalent, that is, there are only Co$^{(4-x)+}$ ions.
The distorted Co-O$_6$ edge-sharing and the layered structures make
t$_{2g}$ manifold split into a$_g$+e$_g^\prime$.
Upon introducing U (there is no agreement
on the best values), both magnetic moments decrease linearly at small U.
At a critical value U$_c$, the moments change discontinuously due to a
first-order transition: the state with similar magnetic moments no long
is a solution, and the new state shows changes in moments just as
expected from charge disproportionation
Co$^{(4-x)+}$$\rightarrow$$x$Co$^{3+}$+$(1-x)$Co$^{4+}$.  This change 
occurs simultaneously with metal-insulator transition 
associated with the Mott transition of the $a_g$ orbital.
Before CD, the $a_g$ band has a similar band center
and width as the doubly degenerate e$_g^{\prime}$ band.

One example of this CD is shown in Fig. \ref{um} for $x=0.5$,
note that it is possible to follow hysteresis within the 
region 3.2 $\leq U\leq$ 3.6 eV and outside this range 
only one of the states, CD or undisproportionated (UD), is obtained.
The two solutions can be observed in another way,
using fixed spin moment (FSM) calculations
within LDA+U in the critical region (bottom panel of Fig. \ref{um}). 
In this method, the energy is calculated versus a constrained value of 
the total moment, and the energy versus
total magnetic moment plot becomes two curves, in which one state or the
other is reached depending on starting point. 
Compared with LDA FSM results by Singh,\cite{singh}
our LDA+U FSM shows two
differences that are worthy of note.
First, at $M=0 \mu_B$, this system is antiferromagnetic (AFM) 
with Co magnetic moment
0.34 $\mu_B$; Singh started from a nonmagnetic state at $M$=0 and 
kept all Co ions identical.
Secondly, at small $M$, the energy vs. total moment plot is nearly
flat, but having two extrema [stable at $M=0.16\mu_B$ and
unstable at $M=0.5\mu_B$].  Due to the miniscule energy barrier, this
small net moment minimum will not be accessible experimentally. 

\begin{figure}[tbp]
\rotatebox{-90}{\resizebox{5.9cm}{5.9cm}{\includegraphics{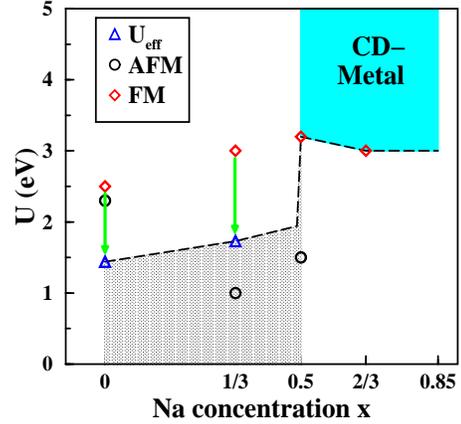}}}
\caption{Effect of Na concentration $x$ on the critical on-site Coulomb 
 repulsion U$_c$ for CD. 
 Allowing antiferromagnetic ordering, U$_c$ drops largely by about 2 eV
 for $x > 0$, while the change is negligible for $x=0$.
 The arrows denote drop of the effect on-site Coulomb repulsion U$^{eff}$
 due to three-band nature for $x < 0.5$ (see text).
 Considering ferromagnetic ordering, comparison with experiments
 suggests a change of U with sharp jump at $x=0.5$, 
 depicted by the shaded region.
 The data are from Ref. \cite{prb04,prb05,4co}.}
\label{ux}
\end{figure}

As shown in Fig. \ref{ux} and described in the caption, 
the critical value U$_c$ for disproportionation depends on
the carrier concentration $x$ and the type of magnetic order. 
In the lower half of the range $x\le 0.5$, U$_c$ 
increases linearly for ferromagnetic (FM) ordering. 
However, allowing AFM ordering, U$_c$($x$) shows
somewhat complicated behavior.
While U$_c$ for both FM and AFM is almost identical at $x=0$,
the value for presumed AFM order decreases sharply for x$\ne$0. 
From Fig. \ref{ux}, U$_c$ for AFM
is smaller by 2 eV than that of FM for $x$=1/3 and $x$=1/2.
(note U$_c$$\sim$ 1 eV for AFM at 
$x=$1/3). [The supercells we used
do not allow for combined disproportionation and AFM ordering for $x=$2/3.]

Furthermore, comparison with experiments suggests that U($x$) is 
significantly dependent on the carrier concentration $x$.
First, no observation of CD nor any significant 
correlation effects at all for $x<0.5$ indicates
that U($x<0.5$) is less than U$_c$.
However, the study of Gunnarsson and coworkers shows that 
in a N-multiband system the effective Coulomb repulsion becomes
U$^{eff}$=U/$\sqrt{N}$.\cite{gunnarsson}
Below U$_c$, the $a_g$ band is only distinguishable from the $t_{2g}$
manifold (in energy) by lying slightly higher, and hosting most of the 
$1-x$ holes. As a result the system at $x<0.5$ maintains 
multiband (three-band) behavior,
resulting in U$_c^{eff}$=U$_c$/$\sqrt{3}$$\sim$ 1.4 eV and 1.7 eV
for $x=0$ and $x=1/3$, respectively.
Since only a tiny band gap ($\sim$ 15 meV) is observed at $x=0.5$,\cite{wang}
U($x=0.5$) must be near the minimum value for CD, i.e. $3.5-4$ eV.
For $x>0.5$, the observed Curie-Weiss susceptibility indicates
$(1-x)$Co$^{4+}$ and $x$Co$^{3+}$ ions, so that the charge is
disproportionated even though the system is a good metal.\cite{mukhamedshin}
Thus, U($x>0.5$) is above 3 eV, perhaps over 4 eV.
We will discuss this $x$=0.5 state elsewhere.  It remains for further
research to decide how accurately LDA+U should be for the band gap, or
whether dynamics of the correlations are important.

\begin{figure}[tbp]
\rotatebox{-90}{\resizebox{6.2cm}{6.9cm}{\includegraphics{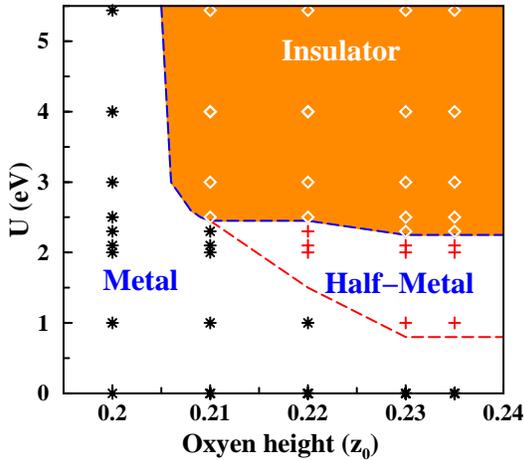}}}
\caption{Phase diagram depending on oxygen height ($z_0$) and U 
 for ferromagnetic ordering at $x=0$. (Energetically, $z_0=0.235$
 is favored.) Lattice constant $c= 4.2509$ \AA~ is used.
 The symbols describe metallic (*), half-metallic (+), 
 and insulating (diamond) states.
 The data are from Ref. \cite{prb05}.}
\label{phase}
\end{figure}

Another issue which is drawing attention in these cobaltates is
whether the change of O height (i.e., Co-O bond-length) is important
as the carrier concentration $x$ increases.
Johannes et al. have calculated the oxygen height $z_0$ by energy minimization,
and concluded that the change has little effect
in the range $0.3\le x \le 0.75$.\cite{johannes}
However, their calculations, and therefore conclusions,
are confined to the local density approximation results and predictions.
We have carried out a study at $x$=0 to address whether
correlation effects described by the LDA+U method are affected by
the oxygen height.
Figure \ref{phase} shows phase diagram in the relevant portion of
the U - $z_0$ plane 
in the relevant range of U and allowing FM ordering of the system.

For $z_0 \ge 0.21$, the system undergoes a metal-insulator
transition,
implying CD, as U is increased.
The critical value U$_c$ depends little on O height.
In fact, $\Delta$U$_c$ is only 0.1 eV between $z_0=0.22$ and
$z_0=0.23$, which is the 
same variation of O height as measured
between $x=0.3$ and $x=0.7$.
This result suggests that the critical value U$_c$ for CD is little affected
by O height in this system.
In addition, it is worthwhile to note that a metal to half-metal
transition
occurs even for very small U before the transition to insulator 
(for example, at U=1 eV for
$z_0=0.23$).

Note however that for a height $x_0$=0.20, no CD is obtained for
U up to 5.5 eV, much different from the behavior for $z_0 \geq 0.21$.
So while this bondlength lies outside the accessible range, it does
show that there is a regime where correlation effects depend very strongly
on the Co-O bond length, and that the system is not so far from that regime.  
Finally, we note that the transitions with varying U 
discussed here should be of special interest for
high pressure research because, for a system
near U$_c$, varying U will be analogous to applying pressure to change
U/W ratio (W is the correlated orbital's bandwidth).


We acknowledge important collaboration with J. Kune\v{s} in the earlier stages of
this work, and have had helpful communications with M. D. Johannes, P. Novak, 
R. T. Scalettar, D. J. Singh, R. R. P. Singh, and J. M. Tarascon.
This work was supported by DOE grant DE-FG03-01ER45876 and DOE's
Computational Materials Science Network.
W. E. P. acknowledges support of the Department of Energy's
Stewardship Science Academic Alliances Program.

\end{document}